\documentclass[review,authoryear,5p,times,twocolumn]{elsarticle}

\usepackage{graphicx}
\usepackage{dcolumn}
\usepackage{bm}
\usepackage{amssymb}
\usepackage{amsfonts}
\usepackage{amsmath}
\usepackage[pdftex]{color}
\usepackage{float}
\usepackage{epsfig}
\usepackage{booktabs} 
\usepackage{array} 
\usepackage{paralist}
\usepackage[english]{babel}
\renewcommand{\r} {\textcolor{red}}
\renewcommand{\b} {\textcolor{blue}}

\journal{International Journal of NonLinear Mechanics}

\begin{document}

\begin{frontmatter}

\title{Temperature--driven volume transition in hydrogels: phase--coexistence
and interface localization}
\author[r1]{E.N.M. Cirillo}
                        \ead{emilio.cirillo@uniroma1.it}

\author[r2]{P. Nardinocchi\corref{cor1}}
\ead{paola.nardinocchi@uniroma1.it}

\author[r3]{G. Sciarra}
\ead{giulio.sciarra@uniroma1.it}

\cortext[cor1]{Corresponding Author: P. Nardinocchi, Dipartimento di Ingegneria Strutturale e Geotecnica, via Eudossiana 18, I-00184 Roma, tel: 0039 06 44585242, fax: 0039 06 4884852.}
 \address[r1]{Dipartimento di Scienze di Base e Applicate per l'Ingegneria,\\
 Sapienza Universit\`a  di Roma, via A.\ Scarpa 16, I--00161, Roma, Italy}
 \address[r2]{Dipartimento di Ingegneria Strutturale e Geotecnica, Universit\`a di Roma ``La Sapienza", via Eudossiana 18, I-00184 Roma, Italy}
\address[r3]{Dipartimento di Ingegneria Chimica Materiali Ambiente, \\
 Sapienza Universit\`a  di Roma, via Eudossiana 18, I-00184 Roma, Italy}
\begin{abstract}
We study volume transition phenomenon in hydrogels within the framework of Flory--Rehner thermodynamic modelling; we show that starting from different models for the Flory parameter different conclusions can be achieved, in terms of admissible coexisting equilibria of the system. In particular, with explicit reference to a one--dimensional problem we establish the ranges of both temperature and traction which allow for the coexistence of a swollen and a shrunk phase. Through consideration of an augmented Flory--Rehner free--energy, which also accounts for the gradient of volume changes, we determine the position of the interface between the coexisting phases, and capture the connection profile between them.
\end{abstract}

\begin{keyword}
hydrogels, phase transition, interface localization.
\end{keyword}
\end{frontmatter}

\section{\label{sec:level1}Introduction}
Since the $'80$, temperature--induced discontinuous phase transitions were largely investigated from an experimental point of view in prolacrylamide gels and in nonionic N--isopropylacrilamide (NIPA) gels, either under free conditions
or uniaxial tension \citep{tanaka1978,tanaka1980,tanaka1984,hirotsu1987,hirotsu1989,suzuki1994,suzuki1997}. 
In particular, in \citep{hirotsu1987},  the results of the experiments were analyzed on the basis of  the Flory--Rehner thermodynamic theory, in which the interactions between polymer and solvent are accounted for by the Flory parameter or polymer--solvent 
interaction parameter which measures the 
dis--affinity between the polymer and the solvent. 
When the Flory parameter is large, solvent molecules are expelled from the gel and 
the gel shrinks, whereas when it is small, 
the gel swells \citep{doi2009}.
In spite of the large amount of data on temperature--dependent volume transition in gels, multiphysics--based models of thermally--driven volume transition in hydrogels have only been recently proposed \citep{ji2006,birgesson2008,duda2010,chester2011, cai2011,ding2013,hong2013,drozdov2014, drozdov2015}.

We make some progress in this topic 
by using an approach based on the Flory--Rehner free energy, 
appropriately extended to penalize gradients of volumetric strain. 
In this way we are able to describe coexistence between swollen and 
shrunk phases and to identify the 
interface position between them. 
In \citep{hong2013}, a similar approach was pursued, even if lacking of any predictions of the spatial distribution of the swollen and the shrunk phases at equilibrium. Our further contribution  consists in discussing and solving the problem with explicit reference to two  different temperature--dependent models of the Flory parameter: the first 
proposed in \citep{hirotsu1989} for NIPA hydrogels  
and the second 
proposed in \citep{afroze2000} for aqueous solutions of uncrosslinked PNIPAM. It is worth noting that in \citep{hong2013}, only this last situation is considered, even if, as we explicitly show through a one--dimensional example, the two models deliver different conclusions.

With reference to a one--dimensional example, we propose an analysis, aimed to identify specific ranges of both temperature and traction which allow  coexistence of the swollen and shrunk phases. The position of the interface is evaluated through an analytical method borrowed from \citep{CIS2012}, and shortly reviewed in the paper, under the limit for the higher order stiffness parameter, accounting for interfacial energy, going to zero. We found that the interface position between the coexisting phases depends on the temperature so that the relative portion of the sample occupied by the shrunk and the swollen phase change with temperature as well. We also verified this prediction through a  numerical calculation.
%
%
\section{Equilibrium theory of gels}
\label{Eqt}
%
Hydrogels are made of long--chain polymers which are cross--linked into a three--dimensional network and permeated by a solvent. From the point of view of continuum mechanics, they can be viewed as soft elastic materials consisting of an elastic matrix swollen with a fluid, and can be thermodynamically characterized through the choice of a free--energy. 
We start considering only isotropic deformation processes, which typically occur under free--swelling conditions; in this case, the state of the system can be described by  the polymer fraction $\phi$, which measures, locally, the 
fraction of volume occupied by the polymer, and by the change in volume $J$, which measures the change in volume from the dry state to the current state of the soft elastic material corresponding to the hydrogel\footnote{\r{If we assume the material to be hyperelastic, the dependence of the free energy only on $J$ is totally general, for isotropic deformations and is not related to the space dimension. The corresponding force is therefore given as the derivative of the energy with respect to $J$.}}. Moreover, we assume that both the volume of the 
solid and liquid component of the hydrogel do not change separately, and the change in volume of the system is a consequence of the variation of the fluid mass content, only; under this assumption, we have that $J=1/\phi$.
With this, the total energy density of the system only
depends on the field $J$, and is parameterised by the temperature $T$ of the hydrogel and the pressure $P$ acting on the solid part of the system. Following \citep{doi2009}, and introducing the relative change in volume between the dry and the current state, 
i.e., $S=J-1>0$, we assume that the above mentioned total energy density $G(S)$ per unit dry volume can be represented as
\begin{equation}
\label{fmvt010}
\begin{array}{rcl}
G(S)&=&
{\displaystyle
 \frac{RT}{\Omega}
 \Big[
     \frac{d}{2N_x}[(S+1)^{\frac{2}{d}}-1]
 \vphantom{\bigg\{_\}}
}
\\
&&
{\displaystyle
 \phantom{
          \frac{RT}{\Omega}
          \Big[
         }
     +
     S
     \log\frac{S}{S+1}+\frac{\chi S}{S+1}
\Big]+PS\b{\,.}
}
\end{array}
\end{equation}
Here, we have: the gas constant $R=8.314$~JK$^{-1}$mol$^{-1}$,
the solvent molar volume $\Omega$, the number $N_x$  of segments in a polymeric chain, the scalar $d$ accounting for the dimension of the environment under consideration, and the Flory parameter 
$\chi$ being a temperature--dependent 
dimensionless parameter (called also the chi parameter), which represents the 
dis--affinity between the polymer and the solvent. 

Let us note that the first addend in the equation (\ref{fmvt010}) can be recognised as
the isotropic elastic energy. 
Indeed, for $d=3$ (three--dimensional context) once introduced the shear modulus
$G=k_\textrm{b}T\,\nu_c$ of the polymer, with $k_\textrm{b}$ the Boltzmann constant and $\nu_c$ the number of 
partial polymeric chains per unit volume, we get
$3G((J^{1/3})^2-1)/2$, 
which is exactly 
the isotropic elastic energy of three--dimensional 
elasticity when only isotropic deformation $J^{1/3}$ are involved (see \citep{JMPS2013}). 
For $d=2$ and $d=1$, analogous considerations can be made to identify 
the first addend with the isotropic elastic contribution: 
for $d=2$, it holds under the assumption of plane deformations $J^{1/2}$; for $d=1$, under the assumption of uniaxial deformation $J$.
\subsection{Modeling the Flory parameter $\chi$}
One of the major assumptions of the Flory--Rehner thermodynamics model is that there is no volume change of the two components on mixing and that solvent and polymer can fit on the sites of the same lattice. It leads to a temperature independent additive constant in the expression of the Flory interaction parameter. Actually, these effects are not fully understood and all deviations from the lattice model, at the basis of the mixing component of Flory--Rehner free energy, are lumped into the interaction parameter $\chi$ which can have non--trivial dependences on polymer fraction and temperature \citep{rubinstein}. Empirically, the temperature dependence of the Flory interaction parameter is written as the sum of two terms: the first is a temperatureÐ independent constant referred as the entropic part of $\chi$; the second depends on temperature and is called the enthalpic part.
With the aim to highlight the differences in phase transition due to the model of $\chi$, we assume that the interaction parameter depends on the volume fraction $\phi$ linearly, according to the model proposed in \citep{hirotsu1989} for NIPA gels and in \citep{afroze2000} for aqueous solutions of uncrosslinked PNIPAM. This choice can be thought of as the first order Taylor expansion of an unknown function of $\phi$:
\begin{equation}
\chi=\chi_{0,T} + \chi_{1,T}\,\phi\,.
\end{equation}
As far as the dependence on the temperature is concerned we consider two different models. The first one, denoted from now on \textit{model a}, prescribes that
\begin{equation}
\chi_{0,T}=A_0 +B_0\,T\quad\textrm{and}\quad
\chi_{1,T}=A_1 +B_1\,T\,,
\end{equation}
with $A_0$, $B_0$, $A_1$, $B_1$ reals; this is the choice which has been demonstrated experimentally to be valid for  PNIPAM \citep{cai2011}  with $A_0 = 12.947$, $B_0 = 0.04496$K$^{-1}$, $A_1 = 17.92$, and $B_1=0.0569$K$^{-1}$1. The second model, denoted from now on \textit{model b}, prescribes
\begin{equation}
\chi_{0,T}=A_0 +\frac{B_0}{T}\quad\textrm{and}\quad
\chi_{1,T}=A_1\,,
\end{equation}
with $A_0 = 2.68294$, $A_1 = 0.305$, and $B_0 = 589.348$K; this is the model proposed in \citep{hirotsu1989} for NIPA hydrogels.
%
\section{Phase coexistence in hydrogels}
\label{s:fwvt}
In this section we set the problem of phase coexistence for a hydrogel described by the energy density \eqref{fmvt010}. We start assuming that a ,homogeneous phase of the system be a global minimum of the energy density \eqref{fmvt010} corresponding to a constant field $S$ and ask for the existence of parameters $T$ and $P$ of the model such that the system is in phase coexistence regime, that is to say, the energy density \eqref{fmvt010} has multiple isolated local minima. 


We split the analysis into two cases: for the interaction parameter $\chi$ depending on the sole temperature $T$ and for the interaction parameter $\chi$ depending on both the temperature $T$ and the volume fraction $\phi$; and discuss the nature of the solutions of the equation $\partial\,G/\partial\,S=0$. We show that for $\chi$ depending on the sole temperature, the existence of multiple coexisting phases is possible only in dimension one $(d = 1)$; whereas for $\chi$ depending on $T$ and $\phi$, it will turn out that coexistence is possible at any dimensions. 
\subsection{Interaction parameter depending on the sole temperature}
\label{s:chicos}
\begin{figure}[h]
\begin{picture}(200,70)(-20,0)
\setlength{\unitlength}{.030cm}
\thinlines
\put(0,0){\vector(0,1){80}}
\put(-5,30){\vector(1,0){80}}
\thicklines
\qbezier(0,30)(50,30)(70,80)
\qbezier[30](0,20)(35,22.5)(70,25)
\qbezier[30](0,20)(35,37)(70,54)
\qbezier[30](0,20)(35,40)(70,70)
\put(73,23){${\scriptscriptstyle S}$}
\put(15,15){${\scriptscriptstyle \chi_T<\chi(N_x)}$}
\put(45,35){${\scriptscriptstyle \chi_T=\chi(N_x)}$}
\put(10,55){${\scriptscriptstyle \chi_T>\chi(N_x)}$}
\thinlines
\put(90,0){\vector(0,1){80}}
\put(85,30){\vector(1,0){80}}
\thicklines
\qbezier(92,5)(92,35)(155,70)
\put(163,23){${\scriptscriptstyle S}$}
\put(91,80){${\scriptscriptstyle L(S)}$}
\thinlines
\put(180,0){\vector(0,1){80}}
\put(175,30){\vector(1,0){80}}
\thicklines
\qbezier(182,5)(182,40)(190,40)
\qbezier(190,40)(195,40)(200,30)
\qbezier(200,30)(205,20)(210,20)
\qbezier(210,20)(220,20)(235,75)
\put(253,23){${\scriptscriptstyle S}$}
\put(181,80){${\scriptscriptstyle L(S)}$}
\end{picture}
\caption{Left panel: qualitative study of equation (\ref{ciccio}) in the case $d=1$;
dotted and solid lines are the graphs of the 
functions on the left and right hand sides of the equation.
Central \& right panels: qualitative 
graphs of the function $L(S)$ introduced below (\ref{fwvt100}),
$\chi_T\b{<}\chi(N_x)$ (center) and 
$\chi_T>\chi(N_x)$ (right).
Note that for $\chi_T\b{=}\chi(N_x)$ 
an horizontal inflection point is present.}
\label{f:qual}
\end{figure}
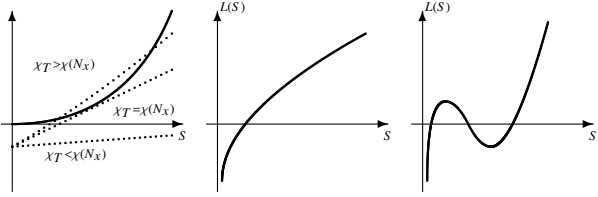

We assume that the interaction parameter $\chi$ does not depend on the volume 
fraction. We then write $\chi=\chi_T$ and the equation  $\partial\,G/\partial\,S=0$ for stationary states as 
\begin{equation}
\label{fwvt100}
     \frac{RT}{\Omega}\Big[\log\frac{S}{S+1}
     +\frac{(S+1)^{\frac{2}{d}-1}}{N_x}
     +\frac{1}{S+1}
     +\frac{\chi_T}{(S+1)^2}\Big]
=
-P\,.
\end{equation}
From now on the left hand side 
in \eqref{fwvt100} will be denoted by $L(S)$. 
In order to understand if equation \eqref{fwvt100} admits one or 
more solutions, few remarks are needed on the behavior of $L(S)$ 
for the order parameter $S$ tending to the admissible 
limits $0$ and $\infty$ as well as on its minima. 
In particular we note 
that
\begin{equation}
\lim_{S\to 0}L(S)=-\infty\,,
\end{equation}
whereas
\begin{equation}
\lim_{S\to\infty}L(S)=\infty,\,\, (1/N_x) (RT/\Omega), \,\, 0\,,
\end{equation}
for $d=1,2,3$, respectively.
Moreover, we have that the equation $\partial L(S)/\partial S=0$ for the stationary 
points of $L$ reads
\begin{equation}
\label{ciccio}
-1+(2\chi_T-1)S=\frac{1}{N_x}\Big(\frac{2}{d}-1\Big)S(S+1)^{2/d+1}\b{\,.}
\end{equation}
A graphical study of the above equation
yields the following: 
for $d=3$ the function $L$ has a single stationary point; 
for $d=2$ the function $L$ has a single stationary point for 
$\chi_T>1/2$ and no stationary point otherwise; 
for $d=1$ there exists a real number
\footnote{
It is possible to give a nice estimate of the number $\chi(N_x)$. 
Indeed, for $d=1$ the equation \eqref{ciccio}
can be rewritten as 
$$
\frac{1}{N_x}S^4
+\frac{3}{N_x}S^3
+\frac{3}{N_x}S^2
+\Big[-(2\chi_T-1)+\frac{1}{N_x}\Big]S
+1
=0
\;\;.
$$
Since, the signs of the coefficient of the above polynomial can exhibit
at most two variations, the number of its positive roots is at most 
equal to two. The condition for having two variations is 
$\chi_T>1/2+1/(2N_x)$. 
Hence we have that $\chi(N_x)>1/2+1/(2N_x)$.
}
such that the function $L$ has two stationary points for 
$\chi_T>\chi(N_x)$, 
one stationary point for $\chi_T=\chi(N_x)$, 
and no stationary point otherwise  (see figure~\ref{f:qual}, left panel). 

From these remarks, it follows 
that the possible qualitative behavior of the function $L$ are 
those depicted in figure~\ref{f:qual} (center and right panels).
The structure of the 
equation \eqref{fwvt100} suggests that it 
can have three solutions only for $d=1$ and $\chi_T>\chi(N_x)$, and
the energy function can have a double well structure only in such a case.
Since for $d=1$ the energy $G(S)$ 
tends to $0$ and to $+\infty$ for 
$S\to0$ and $S\to\infty$, respectively, it follows, 
from the standard Maxwell equal area construction, that it is possible 
to find a value of the pressure/traction $P$ such that 
the energy is characterized by two equally deep wells. 

In conclusion, in the case studied in this section, namely, 
for the interaction parameter $\chi_T$ not depending on 
the volume fraction $\phi$, the hydrogel described by the 
total
energy \eqref{fmvt010} can exhibit two coexisting 
homogeneous swelling states only in the one--dimensional case. 
This possibility is ruled out at larger dimensions.

 \subsection{Free swelling state}
 \label{s:free}
 In the setup of the above Section~\ref{s:chicos}, namely, 
 under the assumption that the interaction parameter $\chi_T$ only depends on $T$, 
 we quickly recall the properties of the equilibrium state 
 under free swelling conditions, i.e, $P=0$ (see also
 the discussion in \citep{doi1996} [Section~3.4]).
 From the result in the previous section, it is immediate that 
 phase coexistence is not possible in the cases $d=2,3$. 
 We prove, here, that even for $d=1$ the phase is unique. 
 For $d=1$ equation \eqref{fwvt100} 
 becomes
 \begin{equation}
 \label{free040}
 (S+1)^2\log\frac{S+1}{S}-\frac{1}{N_x}(S+1)^3-(S+1)
 =\chi_T\b{\,.}
 \end{equation}
 We denote by $g(S)$ the function at the left hand side of the above 
 equation 
 and 
 note that $g(S)\to+\infty$ for $S\to0$, 
 $g(S)\to-\infty$ for $S\to\infty$, and 
 \begin{equation}
 \frac{\partial}{\partial S} g(S)=
 2(S+1)\log\frac{S+1}{S}
 -\frac{S+1}{S}-1
 -\frac{3}{N_x}(S+1)^{2}\,.
 \end{equation}
 Since the sum of the first two terms is smaller than or equal to $1$, 
 we have that $\partial g/\partial S\le 0$ for any $S$. 
 In conclusion, $g$ is a function decreasing monotonically 
 from $+\infty$ to $-\infty$ in the interval for $S\in[0,+\infty)$; 
 hence, the equation \eqref{free040} has a unique not trivial solution
 whatever the value of $\chi_T$ is.
 
 
 In the free swelling case ($P=0$),
 the energy \eqref{fmvt010} is such that 
 $G(S)\to0$ for $S\to0$ and 
 $G(S)\to+\infty$ for $S\to+\infty$.
 Hence, the unique stationary 
 point of $G(S)$ is a minimum.  
 In conclusion in the free swelling case, 
 for any value of $\chi_T$, namely, for any temperature, 
 the system has a unique not trivial homogeneous phase. 
 This result is in agreement with the 
 discussion in \citep{doi1996}. 

\subsection{Interaction parameter depending on both the volume fraction and the temperature}
\label{s:chidep}
In this case, we get 
\begin{equation}
\label{chidep020}
\begin{array}{rcl}
G(S)
&=&
{\displaystyle
\frac{RT}{\Omega}
\Big[
     \frac{d}{2N_x}((S+1)^{2/d}-1)
}\\
&&
{\displaystyle
 \phantom{ \frac{RT}{\Omega} \Big[ }
   + S
     \log\frac{S}{S+1}
+\Big(\chi_{0,T}
\vphantom{\bigg\{_\}}
}
\\
     &&
{\displaystyle
 \phantom{ \frac{RT}{\Omega} \Big[ }
 +\chi_{1,T}\frac{1}{S+1}\Big)\frac{S}{S+1}
\Big]
     +PS\b{\,.}
}
\end{array}
\end{equation}

Our original problem can be now rephrased as follows:
are there any values of the thermodynamic parameters 
$T$ and $P$ which determine two different swelling states of the system? 
In other terms, we look for values of $T$ and $P$ such that 
the energy $G(S)$ has a double well 
graph with equally deep wells. 
For $d=2,3$, it is possible to get a double well energy.
However, we focus on the case $d=1$, 
being dimension one more suitable for setting and solving the  interface location problem, through a technique already proposed in different contexts by some of the 
authors \citep{CIS2009,CIS2010,CIS2011,CIS2012,CIS2013}. 


Firstly, we note that $S>0$ and 
$$
\lim_{S\to0}G(S)=0
\;\;\;\textrm{ and }\;\;\;
\lim_{S\to\infty}G(S)
=
+\infty\b{\,.}
$$
Then, since the number of stationary points of the 
total
energy is determined by its first partial derivative 
computed with respect to $S$, we evaluate
\begin{eqnarray}
\label{chidep040}
\frac{\partial }{\partial S}G(S)
&=&
\frac{RT}{\Omega}
\Big[
\frac{1}{N_x}(S+1)
+\chi_{0,T}\frac{1}{(S+1)^2}\\
&&+\chi_{1,T}\frac{1-S}{(S+1)^3}
+\log\frac{S}{S+1}
+\frac{1}{S+1}
\Big]
+P\nonumber
\end{eqnarray}
and 
$$
\lim_{S\to0}\frac{\partial}{\partial S}G(S)
=-\infty
\;\;\;\textrm{ and }\;\;\;
\lim_{S\to\infty}\frac{\partial}{\partial S}G(S)
=
+\infty \b{.}
$$
The equation determining the stationary points 
$\partial G(S)/\partial S=0$
can be rewritten as
\begin{equation}
\label{chidep050}
L(S)=-\, P\b{\,,}
\end{equation}
where 
\begin{eqnarray}
\label{chidep060}
L(S)
&=&\frac{RT}{\Omega}\Big[
\frac{1}{N_x}(S+1)
+\chi_{0,T}\frac{1}{(S+1)^2}\nonumber\\
&+&\chi_{1,T}\frac{1-S}{(S+1)^3}
+\log\frac{S}{S+1}
+\frac{1}{S+1}\Big]
\end{eqnarray}
defines the traction exerted on the hydrogel.
In order to establish the number of solutions 
of the equation \eqref{chidep050}, we look at the  graph of the function $L(S)$. To do it, we note that
\begin{equation}
\label{chidep070}
\lim_{S\to0}L(S)=-\infty
\;\;\;\textrm{ and }\;\;\;
\lim_{S\to\infty}L(S)=+\infty
\end{equation}
and
\begin{eqnarray}
\label{chidep080}
\frac{\partial}{\partial S} L(S)
&=&
\frac{RT}{\Omega}\frac{1}{S(S+1)^4}
\Big[
     \frac{1}{N_x}S^5
     +\frac{4}{N_x}S^4
     +\frac{6}{N_x}S^3\nonumber\\
&&+\,\Big(\frac{4}{N_x}+1-2\chi_{0,T}+2\chi_{1,T}\Big)S^2
\vphantom{\bigg\{_\}} 
\\
&&
     +\,\Big(\frac{1}{N_x}+2-2\chi_{0,T}-4\chi_{1,T}\Big)S
     +1
\Big]
\nonumber
\end{eqnarray}
The numerator of $\partial L/\partial S$ is a fifth order polynomial
and its number of positive zeros can be estimated by means of the 
Descartes rule: first note that four of the six coefficients are positive 
and the two remaining are the first and the 
second order ones. We then have that,
assuming that all the coefficients of the polynomial are 
different from zero, their sign can exhibit either zero or two variations. 
This implies that $\partial L/\partial S$ has either zero or two positive real zeros. 

Hence, recalling \eqref{chidep070}, the graph of the function 
$L(S)$ can be either monotonic 
or kinky (as we have already illustrated in 
figure~\ref{f:qual}, center and right panel, in a different case)).
The first case corresponds to absence of coexisting 
phases, whereas the 
second  to the existence of two different phases. 
It is worth noting that the presence of two coexisting phase is possible only if 
the pressure is chosen properly via the Maxwell equal area rule. 

\begin{figure}
\begin{picture}(200,150)(-10,0)
\put(40,0)
{
\resizebox{5cm}{!}{\rotatebox{0}{\includegraphics{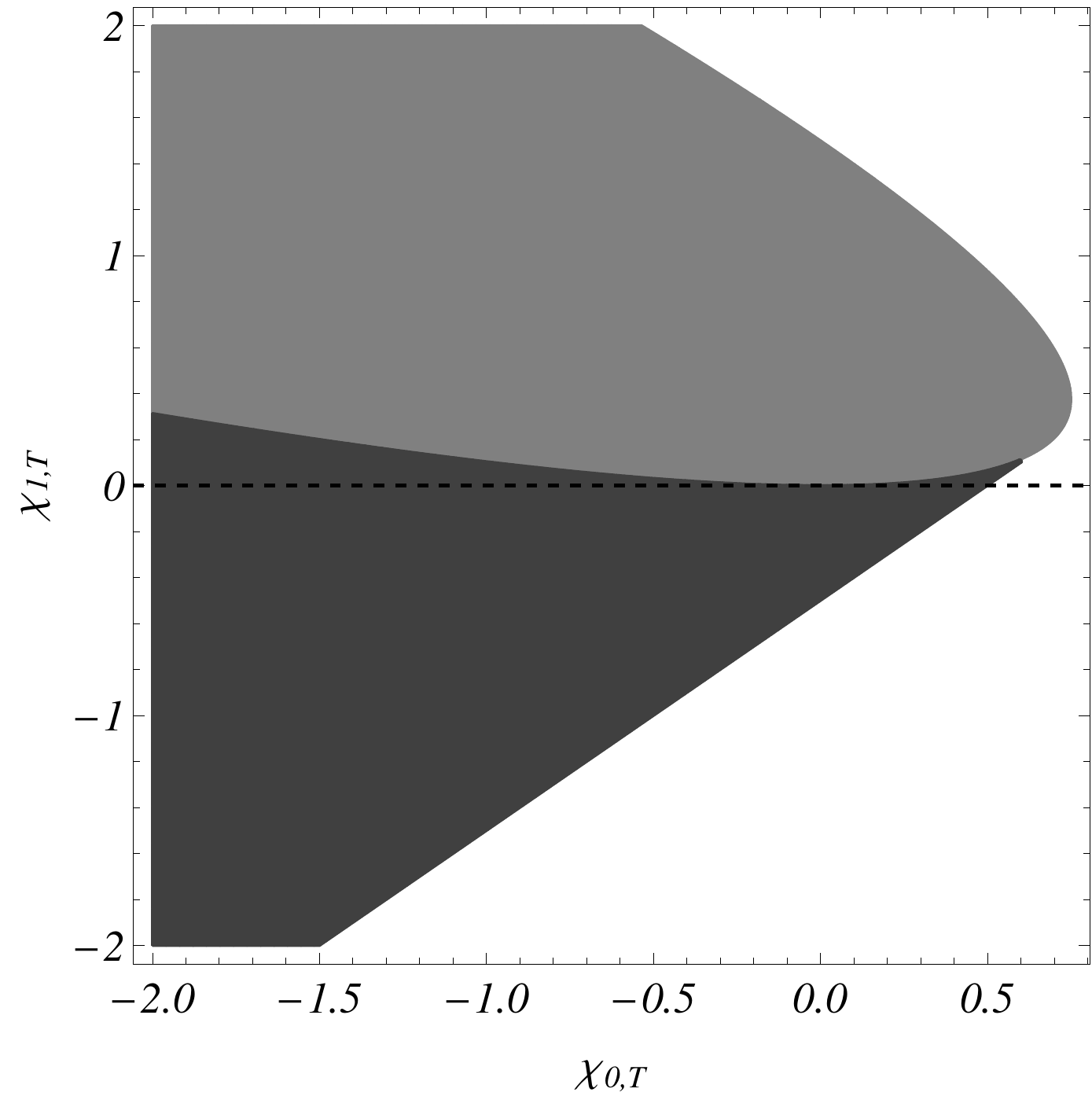}}}
}
\end{picture}
\caption{The light gray region is a graphical representation of the solutions 
of the inequality \eqref{chidep160}. 
The dark gray region is a graphical representation of the solutions 
of the inequality \eqref{chidep170}. Note that the results depicted in this 
figure are consistent with those discussed in Section~\ref{s:chicos}:
indeed, for $\chi_{1,T}=0$ the interaction parameter does not depend 
on the volume fraction, and the graph above is consistent with 
the fact that the system admits coexisting phases provided $\chi_{0,T}$ 
is sufficiently large. 
}
\label{f:chidep010}
\end{figure}

Now the natural question is: for what values of $\chi_{0,T}$ and 
$\chi_{1,T}$ is
the graph kinky? This question is not easy to answer, since 
studying the possibility of a quintic polynomial to have real 
roots is not trivial; we propose a nice estimate. 
With simple algebra we can write the derivative of the function 
$L(S)$ as

\begin{equation}
\begin{array}{rcl}
\label{chidep140}
{\displaystyle 
\frac{\partial}{\partial S} L(S)
}
&=&
{\displaystyle 
\frac{RT}{\Omega}\Big\{
\frac{1}{N_x}
+\frac{1}{S(S+1)^4}
\big[S^2(1-2\chi_{0,T}
}
\\
&&
\phantom{\frac{RT}{\Omega}\Big[}
{\displaystyle 
+2\chi_{1,T})
}\\
&&
{\displaystyle 
\phantom{\frac{RT}{\Omega}\Big[}
 +2S(1-\chi_{0,T}-2\chi_{1,T})+1\big]\Big\}\b{\,,}
}
\end{array}
\end{equation}
and note that,
a sufficient condition for $\partial L/\partial S$ to be strictly positive 
is that the discriminant of the second order polynomial 
appearing in the above equation is negative. Indeed, in such a case 
it would be also ensured that the first coefficient 
$1-2\chi_{0,T}+2\chi_{1,T}$ is positive, so that the second degree polynomial 
is positive defined. Hence, we find the sufficient 
condition
\begin{equation}
\label{chidep160}
(1-\chi_{0,T}-2\chi_{1,T})^2-(1-2\chi_{0,T}+2\chi_{1,T})<0\b{\,,}
\end{equation}
which has the nice geometrical interpretation on the plane 
$\chi_{0,T}$--$\chi_{1,T}$ depicted in 
figure~\ref{f:chidep010} (light gray region).

Another condition ensuring the derivative $\partial L/\partial S$ to be positive 
for $S>0$ 
is that the discriminant is positive but both the 
coefficient of $S^2$ and $S$ are positive too. Indeed, in such a case, 
by the Descartes'rule, it follows immediately that the 
two roots of the second order polynomial are both negative. Thus, 
we have the inequalities 
\begin{equation}
\label{chidep170}
\left\{
\begin{array}{l}
(1-\chi_{0,T}-2\chi_{1,T})^2-(1-2\chi_{0,T}+2\chi_{1,T})>0
\\
1-\chi_{0,T}-2\chi_{1,T}>0
\\
1-2\chi_{0,T}+2\chi_{1,T}>0
\end{array}
\right.
\end{equation}
which has the nice geometrical interpretation on the plane 
$\chi_{0,T}$--$\chi_{1,T}$ depicted in 
figure~\ref{f:chidep010}  (dark gray region).

We remark that the conditions we found are sufficient, but not necessary, 
to rule out the existence of coexisting phases. Indeed, even 
if the second order polynomial in \eqref{chidep140}
were negative for some values of $S$, it could happen that, 
due to the $1/N_x$ additive term, the derivative $\partial L/\partial S$ is
positive for any $S$. 
Hence, the region with absence of multiple coexisting 
phases could be larger than
the one depicted in figure~\ref{f:chidep010}.
We also note that such a condition becomes sharper and sharper
when $N_x$ is chosen larger and larger. 

\begin{figure*}
\begin{picture}(200,70)(-5,0)
\put(0,0)
{
\resizebox{3.84375 cm}{!}{\rotatebox{0}{\includegraphics{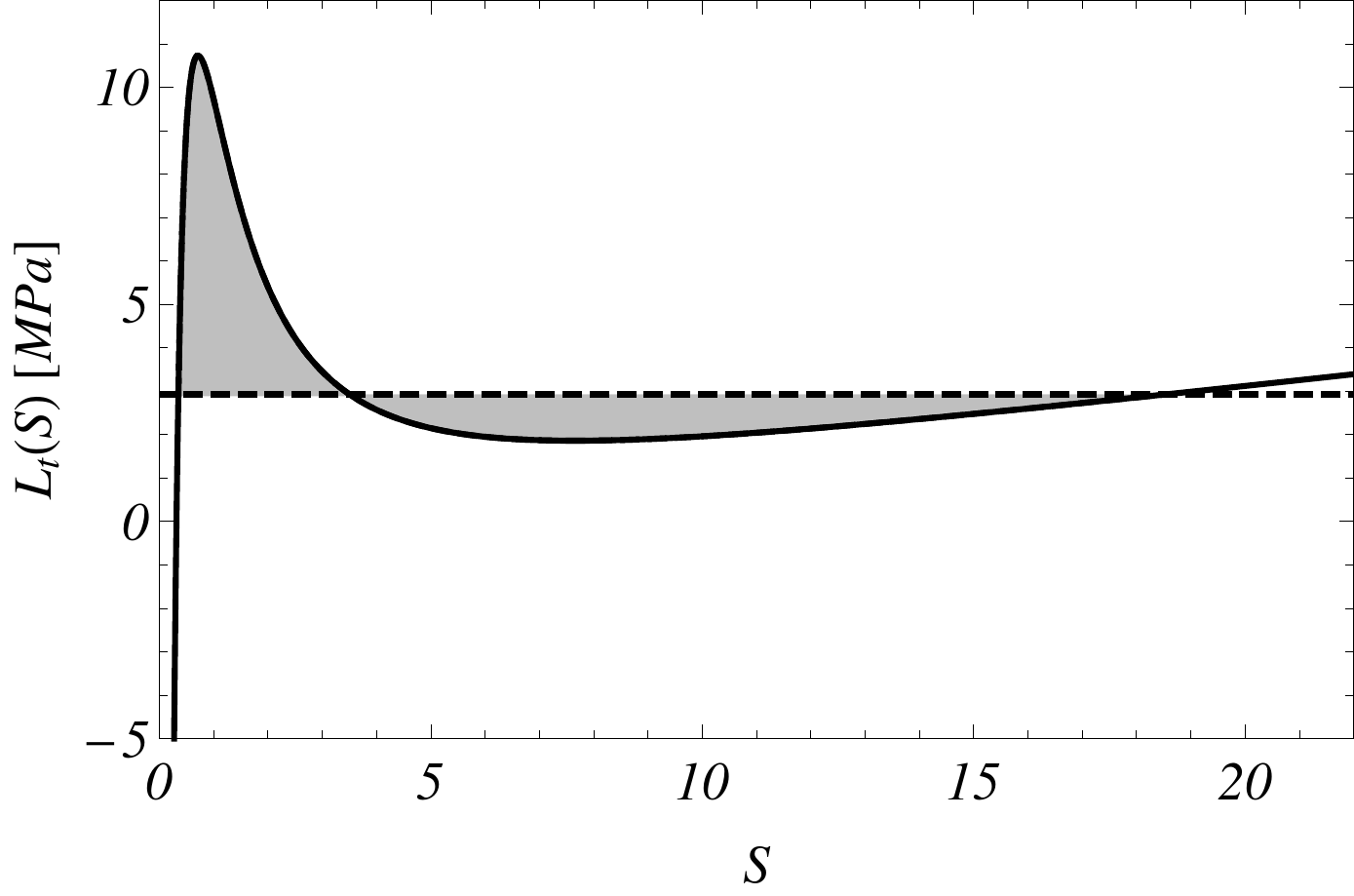}}}
}
\put(125,0)
{
\resizebox{3.9375 cm}{!}{\rotatebox{0}{\includegraphics{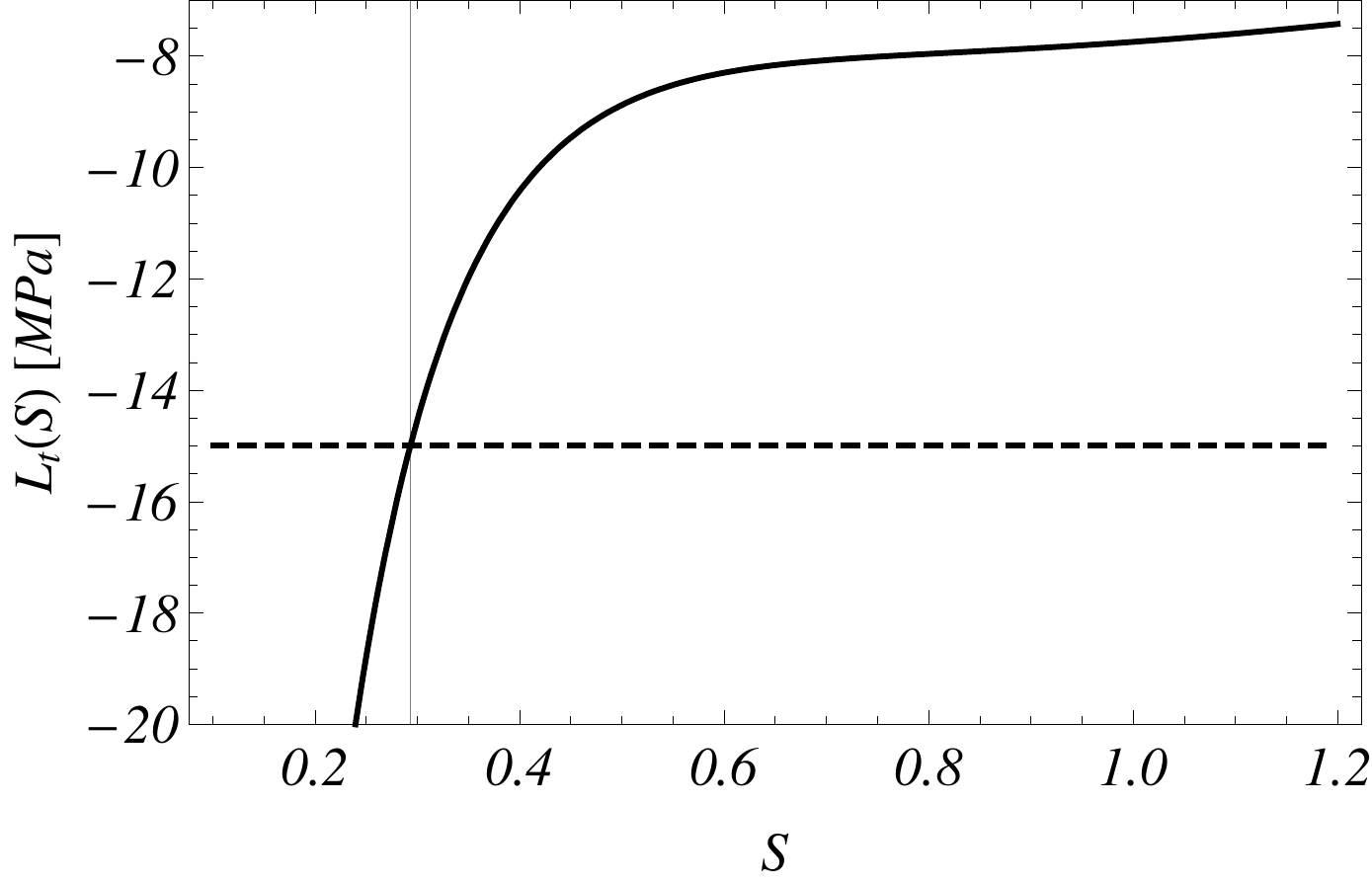}}}
}
\put(250,0)
{
\resizebox{4.125 cm}{!}{\rotatebox{0}{\includegraphics{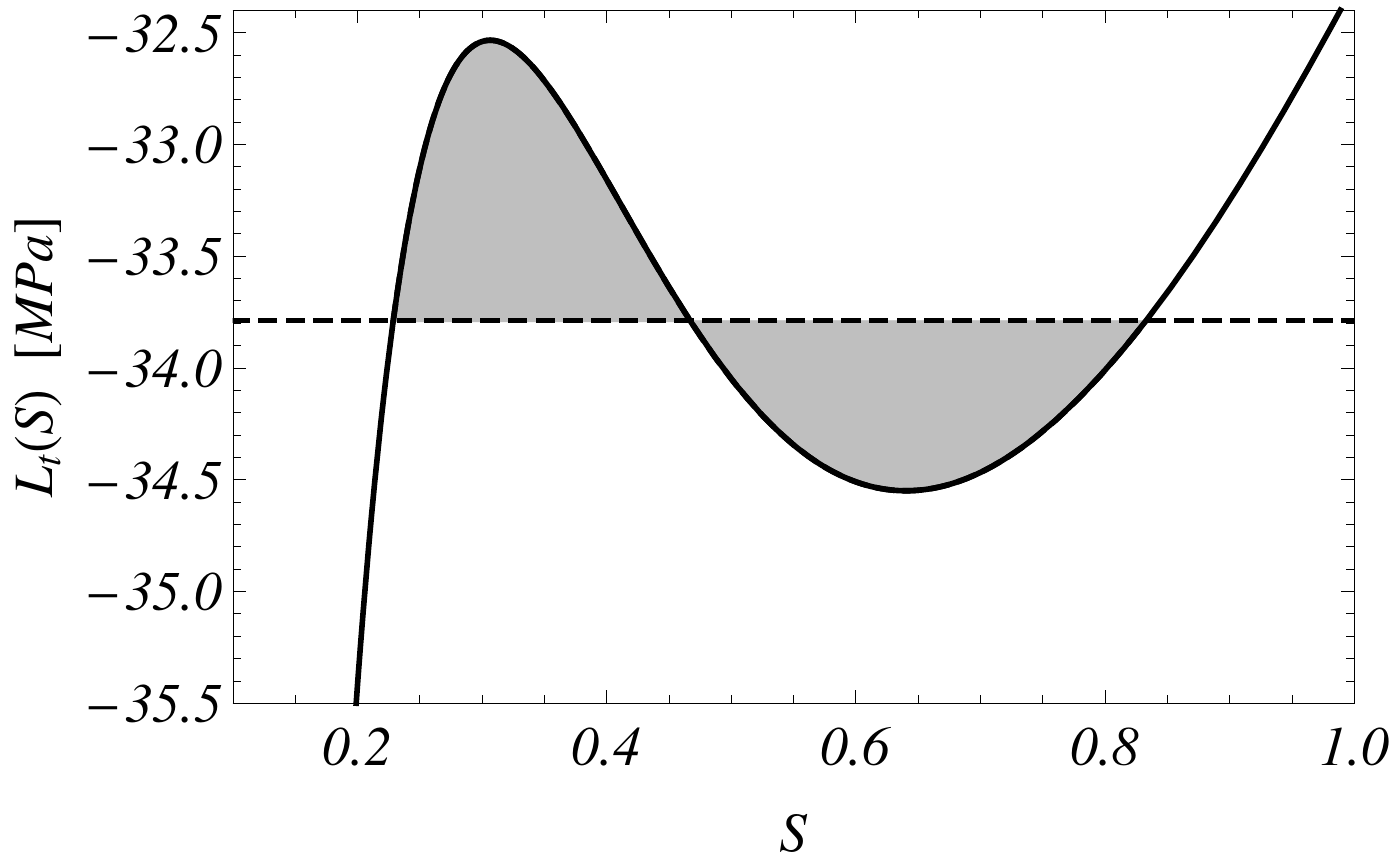}}}
}
\put(390,0)
{
\resizebox{3.9375 cm}{!}{\rotatebox{0}{\includegraphics{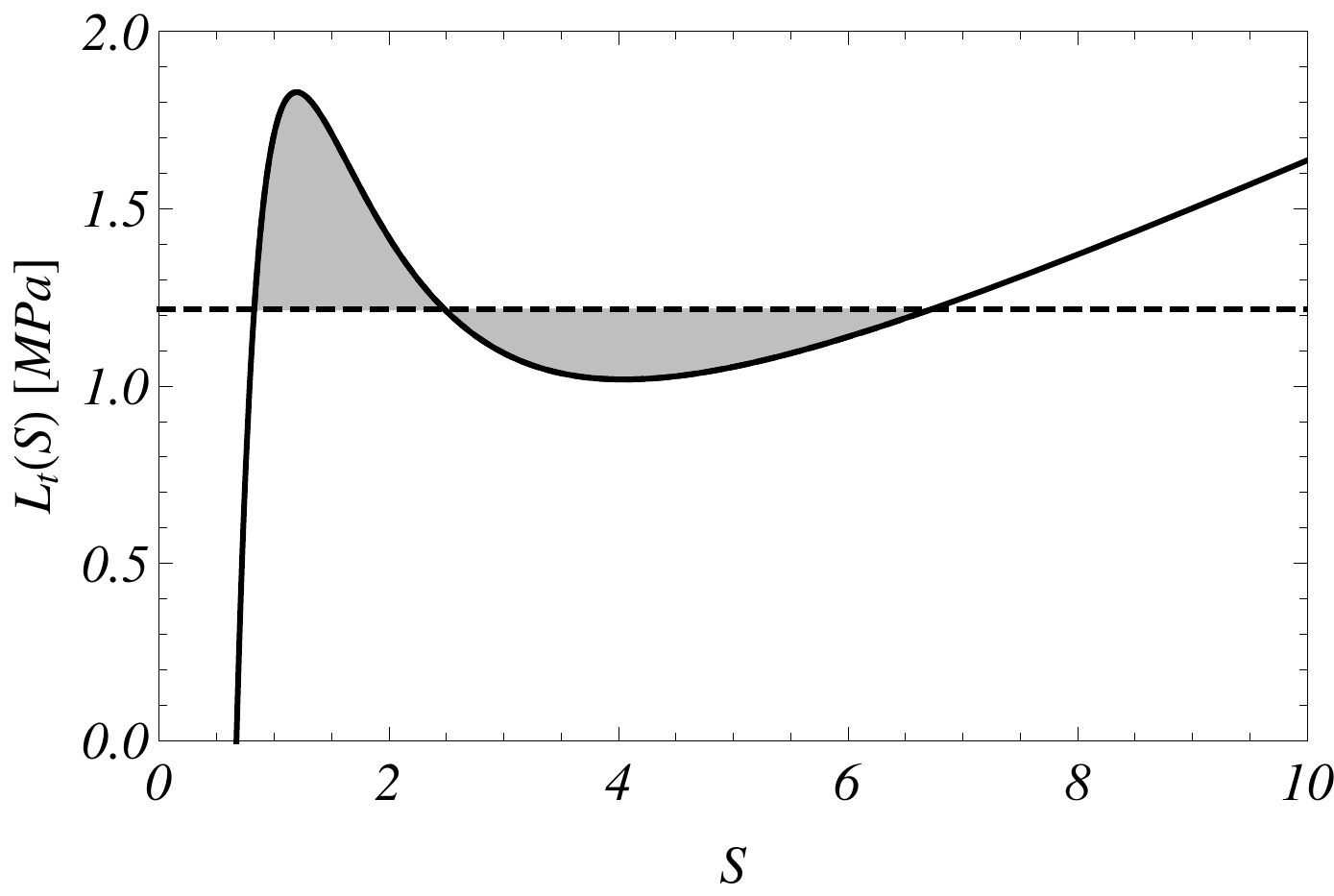}}}
}
\end{picture}
\caption{Graphs of the traction $L(S)$ \eqref{chidep060}. First three
panels with Flory parameter temperature dependence 
as in \textit{model a} and, from the left to the right, 
$T=283$, $300$, $311$~K and 
$P=33.79$, $15.00$, $-2.93$~MPa. Fourth panel with Flory parameter temperature dependence 
as in  \textit{model b}, $T=315$~K and $P=-1.22$~MPa.}
\label{f:chidep030}
\end{figure*}
\section{Temperature driven phase transition in PNIPAM}
\label{s:tdri}
In this section we study the possibility of a temperature driven phase transition in PNIPAM in the framework of the one--dimensional model discussed in Section \ref{s:chidep}. We start assuming it holds \textit{model a} for the Flory interaction parameter.
Under this assumption condition \eqref{chidep160} reads
\begin{eqnarray}
\label{tdri010}
&&(B_0+2B_1)^2T^2
+2[(A_0+2A_1)(B_0+2B_1)
\nonumber\\
&&
\phantom{mmmmmm}
-3B_1]T
+(A_0+2A_1)^2-6A_1<0
\end{eqnarray}
whereas 
conditions \eqref{chidep170} can be rephrased as follows
\begin{equation}
\label{tdri020}
\left\{
\begin{array}{l}
(B_0+2B_1)^2T^2
+2[(A_0+2A_1)(B_0+2B_1)
\\
\phantom{mmmmmi}
-3B_1]T
+(A_0+2A_1)^2-6A_1>0
\\
1-(A_0+2A_1)-(B_0+2B_1)T>0
\\
1-2(A_0+A_1)+2(B_0+B_1)T>0
\\
\end{array}
\right.
\end{equation}

In the PNIPAM case \citep{cai2011}, we
have that \eqref{tdri010} is satisfied 
for $291.198$~K$<T<301.868$~K while \eqref{tdri020} is never satisfied.
This means that the system can exhibit (depending on the 
pressure) two coexisting phases 
for $T<T_{\textrm{low}}<291.198$~K and 
$T>T_{\textrm{high}}>301.868$~K.
We recall that the numerical estimate for $T_{\textrm{low}}$ and 
$T_{\textrm{high}}$ are sharper and sharper provided $N_x$ is larger 
and larger.

This behavior is illustrated in in the first three panels of figure~\ref{f:chidep030}.
The Maxwell construction is illustrated 
at $T=283$~K and $T=311$~K 
in order to find the value of the coexisting pressure, namely, 
the pressure such that the two gray areas in figure~\ref{f:chidep030},
or analogously the two local minima of the energy, are equal. 
Conversely at $T=300$~K the energy has a single minimum, for any 
value of the pressure, so that a unique phase is observed.

If the \textit{model b} of the Flory interaction parameter is assumed, condition  \eqref{chidep160} yields
\begin{equation}
\label{tdri060}
\begin{array}{l}
[-1+2A_0-2A_1+(-1+A_0+2A_1)^2]T^2
\vphantom{\Big\{}
\\
\phantom{mmmmmmmmm}
+2B_0(A_0+2A_1)T
+B_0^2<0
\end{array}
\end{equation}
whereas 
conditions \eqref{chidep170} read
\begin{equation}
\label{tdri070}
\left\{
\begin{array}{l}
[-1+2A_0-2A_1+(-1+A_0+2A_1)^2]T^2
\\
\phantom{mmmmmmmm}
+2B_0(A_0+2A_1)T
+B_0^2>0
\\
-B_0+(1-A_0-2A_1)T>0
\\
-2B_0+(1-2A_0+2A_1)T>0
\\
\end{array}
\right.
\end{equation}
We
have that \eqref{tdri060} is satisfied 
for $126.869$~K$<T<303.825$~K while \eqref{tdri070} is satisfied
for $T<126.869$~K.
This means that the system can exhibit (depending on the 
pressure) two coexisting phases 
for $T>T_{\textrm{high}}>303.825$~K.
This behavior is illustrated in figure~\ref{f:chidep030} (fourth panel).
In the graph the Maxwell construction is illustrated at $T=315$~K
and the coexistence pressure is found.

We remark that the estimate for the $T_{\textrm{high}}$ 
coexistence temperature is very similar in the two theories. 
The main difference is in the fact that with the \textit{model a} phase coexistence is possible even at low 
temperature ($T<T_{\textrm{low}}$), while with the \textit{model b} coexistence is possible only at high temperatures. 

\begin{figure}[h]
\centering
\includegraphics[width=0.35\textwidth]{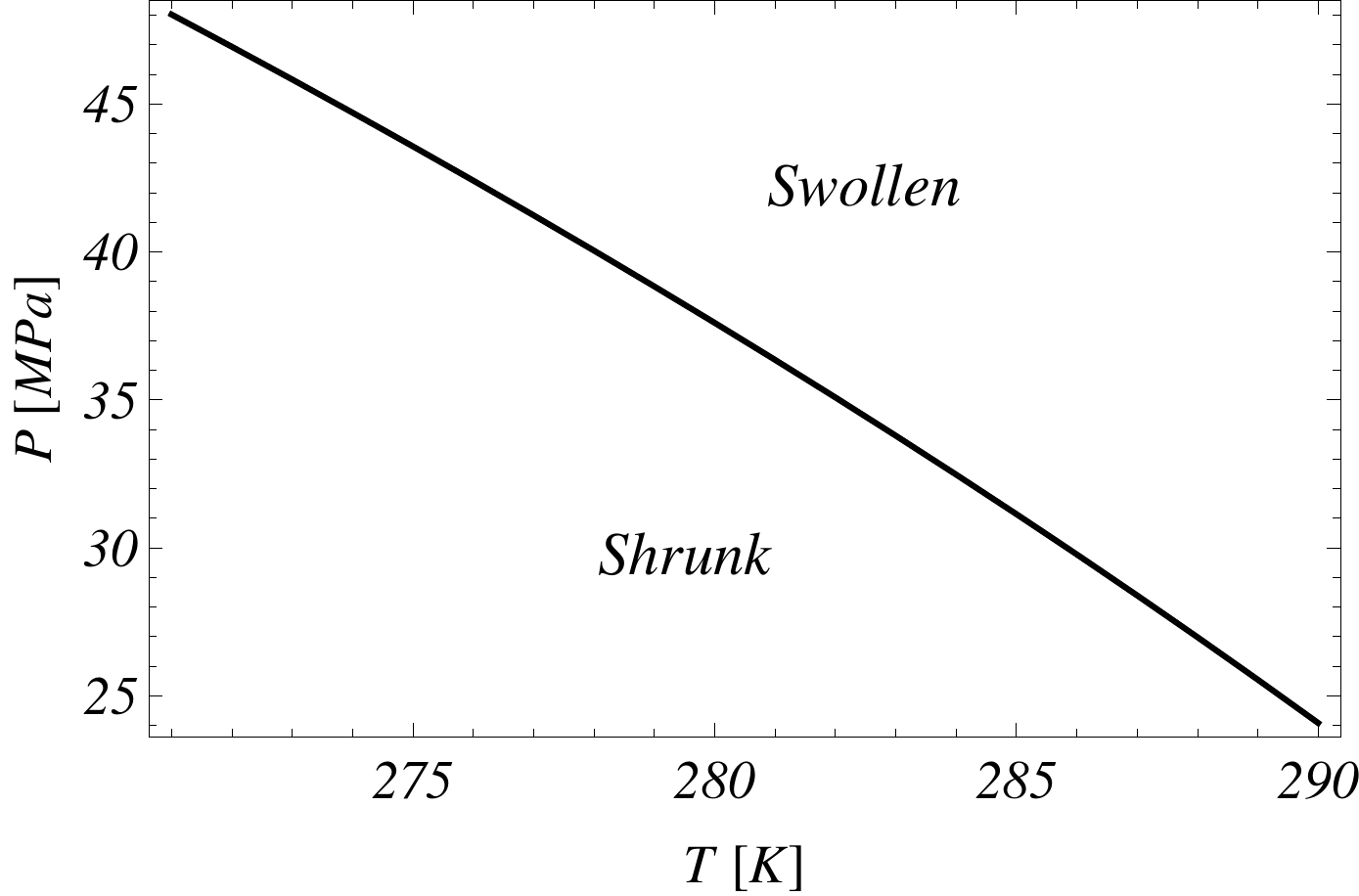}
\includegraphics[width=0.35\textwidth]{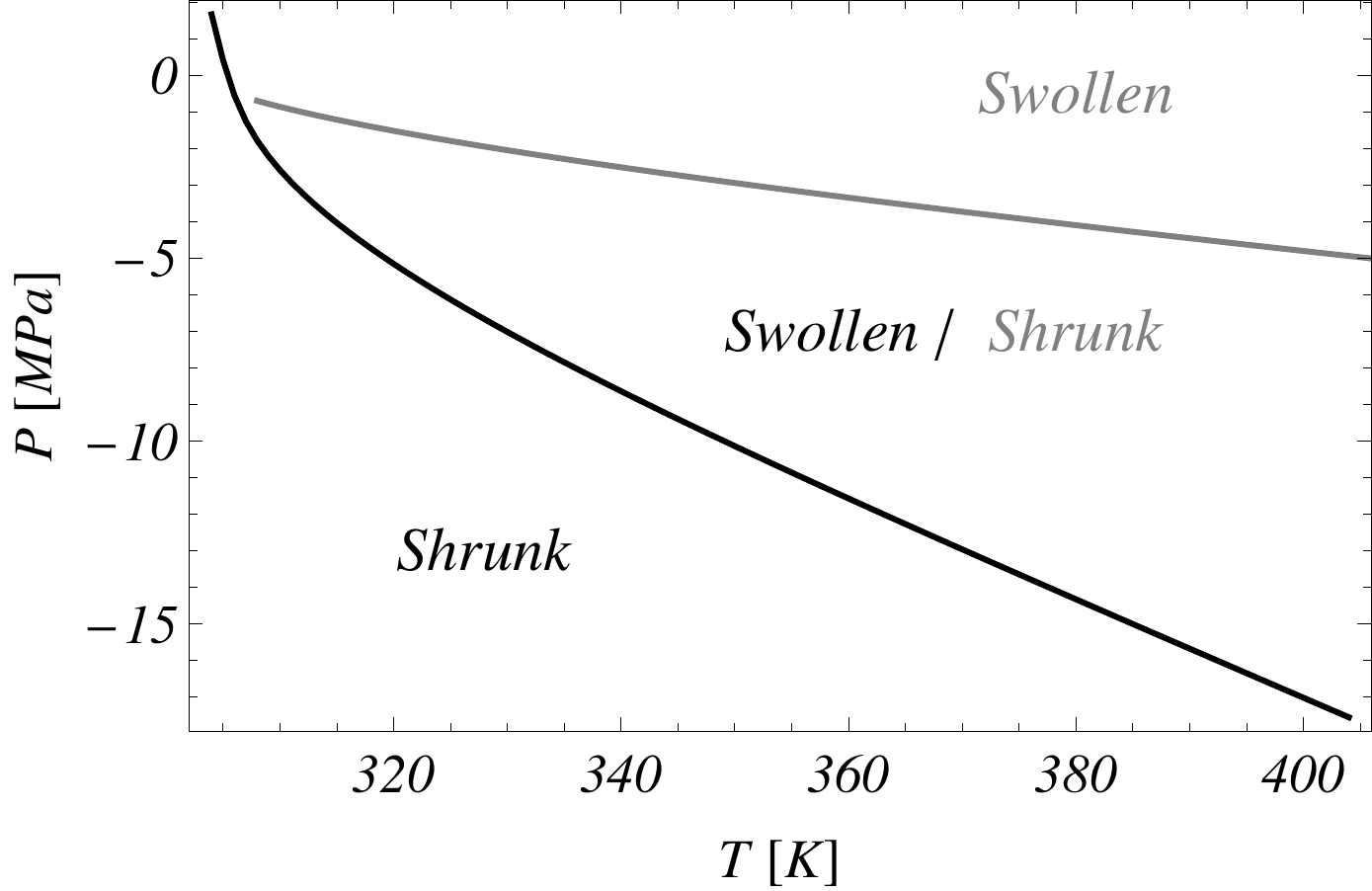}
\caption{Coexistence pressures vs. temperature. Top: results for the \textit{model a} at $T<T_{\textrm{low}}$
are plotted. Bottom: results for the \textit{model a} (black) and for the \textit{model b} (gray) at $T>T_{\textrm{high}}$
are plotted.}
\label{f:chidep060}
\end{figure}


In the following we discuss the physical properties of the system in presence 
of two coexisting phases, 
the \textit{shrunk} and 
the \textit{swollen} 
one, which  
are the admissible equilibria 
with lower and higher volume change.

\begin{figure}[h]
\centering
\includegraphics[width=0.35\textwidth]{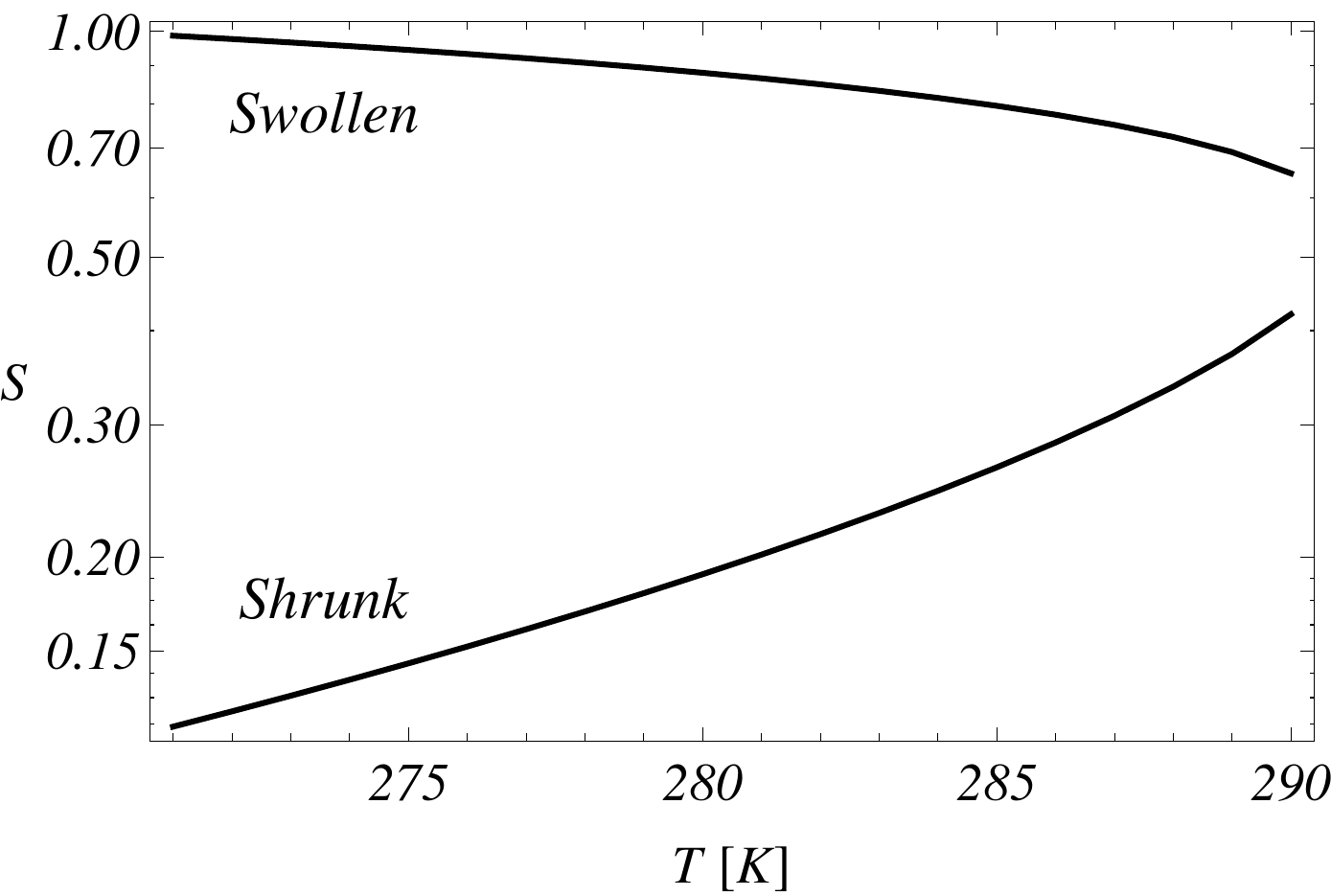}
\includegraphics[width=0.35\textwidth]{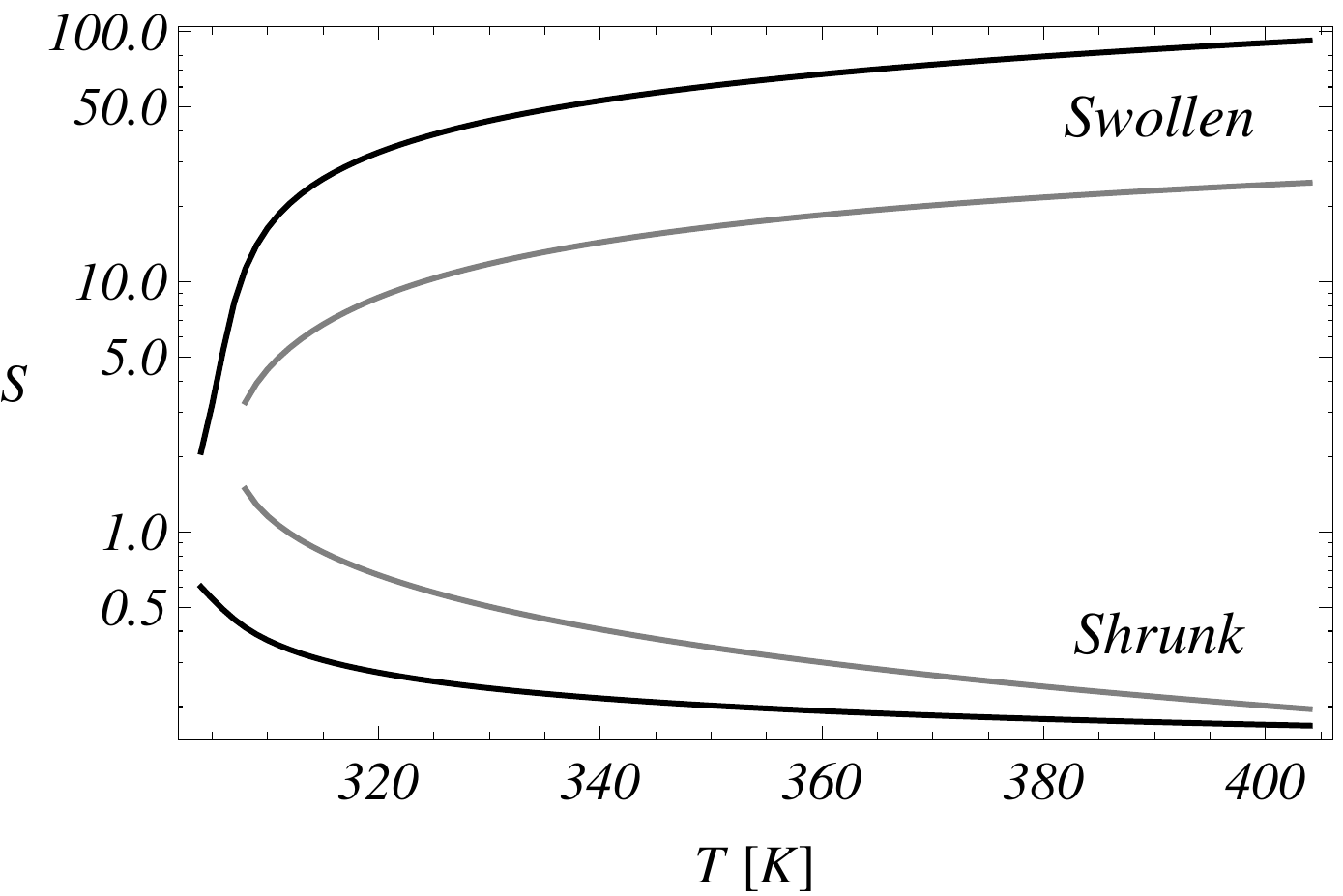}
\caption{Coexisting phases vs. temperature. Top: results for the \textit{model a} at $T<T_{\textrm{low}}$
are plotted. Bottom: results for the \textit{model a} (black) and for the \textit{model b} (gray) at $T>T_{\textrm{high}}$
are plotted.}
\label{f:chidep070}
\end{figure}

In figure~\ref{f:chidep060} we plot the coexistence pressure versus the temperature. On the top we consider the \textit{model a} at $T<T_{\textrm{low}}$
and note that the coexistence between the two phases occurs under compression. 
To the best of our knowledge the possibility of attaining coexistence in such regime has never been
discussed in the related literature. On the bottom we consider both \textit{model  a} and \textit{model b} at $T>T_{\textrm{high}}$, where we stress 
that the two models differ in the value of $T_{\textrm{high}}$ (see the discussion below \eqref{tdri020} and \eqref{tdri070}). The coexistence between the two phases occurs
here under tension for almost all the considered values of temperature, but for the \textit{model a} at temperatures close to $T_{\textrm{high}}$,
where coexistence occurs under compression. It is worth to notice that in the $T$--$P$ plane there exist a region, the one between the two curves, where
the two models predict different phases for the system.

In figure~\ref{f:chidep070} we plot the coexisting phases versus the temperature. On the top, we consider the \textit{model a} at $T<T_{\textrm{low}}$
and, on the bottom, both \textit{model a} and \textit{model b} at $T>T_{\textrm{high}}$. It is worth to notice that the volume changes associated to the shrunk and the swollen
phase are quite similar when $T<T_{\textrm{low}}$, while they significantly differ one from the other at $T>T_{\textrm{high}}$. Increasing the absolute difference between the current and the 
limit temperature, 
either $T_{\textrm{low}}$ or $T_{\textrm{high}}$, 
the shrunk and the swollen phase tend to separate from each other. 
\section{Interface location}
\label{s:secondo}
Gradient theories are suitable to be 
developed for modeling stress/strain concentration due, 
for instance, to the presence of geometrical singularities (crack
propagation in fracture mechanics) or phase transitions
as in the case of wetting.  To the best of our knowledge, gradient theories have  rarely been formulated to infer interface location when temperature--driven volume transition occurs in hydrogels. Some recent results within the framework of phase--field models have been obtained in \cite{hong2013}.

Within the one--dimensional framework introduced above, the gradient theory will be used as a tool to capture the position of the 
interface between two coexisting phases differing in the degree of swelling (shrunk and swollen phases).

We consider the finite interval $[0,1]$ and introduce the Landau energy functional
\begin{equation}\label{enhG}
\mathcal F[S]=\int_0^1 \left[ G(S)+ \frac{\kappa}{2}(S')^2 \right] \mathrm{d}x
\end{equation}
with the standard free--energy component $G(S)$, defined as in the equation (\ref{chidep020}), and $\kappa>0$ an appropriate higher order stiffness; the prime denotes the derivative with respect to the space variable. We look for the equilibrium profile $S(x)$ by performing a standard variational computation and assuming either Dirichlet or homogeneous Neumann boundary conditions. We therefore get the following Euler--Lagrange equation 
\begin{equation}\label{Euler-Lagrange}
\kappa S''  = \frac{\partial G}{\partial S}.
\end{equation}
We consider now Dirichlet boundary conditions corresponding to the shrunk ($S_{\mathrm{sh}}$) and the swollen  ($S_{\mathrm{sw}}$) phases for the values of $P$ and $T$ ensuring coexistence. In other words, we solve the problem \eqref{Euler-Lagrange} with the 
boundary conditions 
\begin{equation}
\label{bound}
S(0)=S_{\mathrm{sh}} 
\;\;\textrm{ and }\;\;
S(1)=S_{\mathrm{sw}}
\;\;. 
\end{equation}

By exploiting the one--dimensionality of the model a phase space 
analysis proves that 
the problem \eqref{Euler-Lagrange} 
endowed with the above mentioned 
boundary conditions has a unique solution implicitly given by the integral 
\begin{equation}
\label{dir01}
\int_{S_{\mathrm{sh}}}^S\frac{\sqrt{\kappa}\,\mathrm{d}s}{\sqrt{2[E_{\kappa}
+G(s)]}}=x
\end{equation}
where for any $\kappa>0$ we have defined implicitly  $E_{\kappa}$
by the equation 
\begin{equation}
\label{dir02}
\int_{S_{\mathrm{sh}}}^{S_{\mathrm{sw}}}\frac{\sqrt{\kappa}\,\mathrm{d}s}{\sqrt{2[E_{\kappa}+G(s)]}}=1
\end{equation}
namely, the integral in (\ref{dir01}) with $S=S_{\mathrm{sw}}$ and $x=1$. 

The profile given by \eqref{dir01} is a connection between the two phases 
and, in the limit $\kappa\rightarrow 0$, 
presents a flat interface localized at 
\begin{equation}
\label{interfaccia}
x_I=\frac{\sqrt{\partial^2 G(S_{\mathrm{sw}})/\partial S^2}}
              {\sqrt{\partial^2 G(S_{\mathrm{sh}})/\partial S^2}+\sqrt{\partial^2 G(S_{\mathrm{sw}})/\partial S^2}}
\end{equation}
see (Theorem~2) in \citep{CIS2012}. 
We note that, if the energy $G$ had equal second derivatives at 
the phases, 
the interface would fall in the middle point of the interval. We underline that in our case the energy \eqref{fmvt010} is not symmetric and therefore the interface position will change depending on the temperature $T$.

\begin{figure*}[t]
\begin{picture}(200,120)(-5,0)
\put(0,0)
{
\resizebox{5.75cm}{!}{\rotatebox{0}{\includegraphics{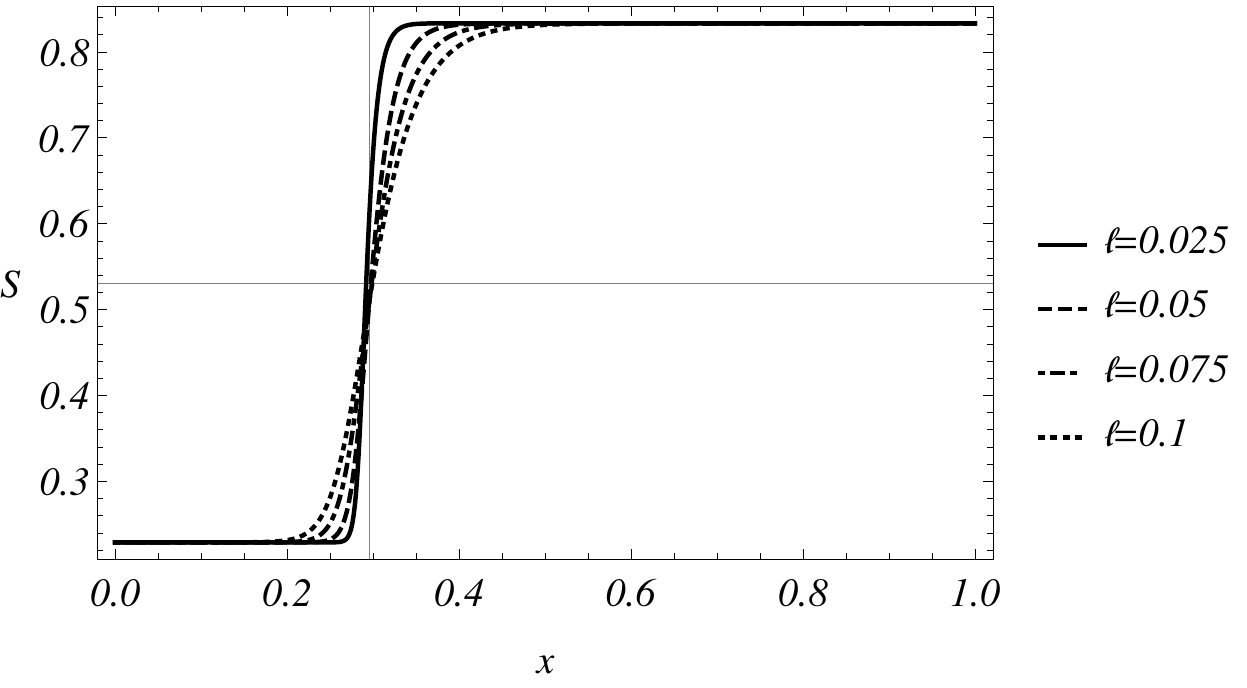}}}
}
\put(170,0)
{
\resizebox{5.75cm}{!}{\rotatebox{0}{\includegraphics{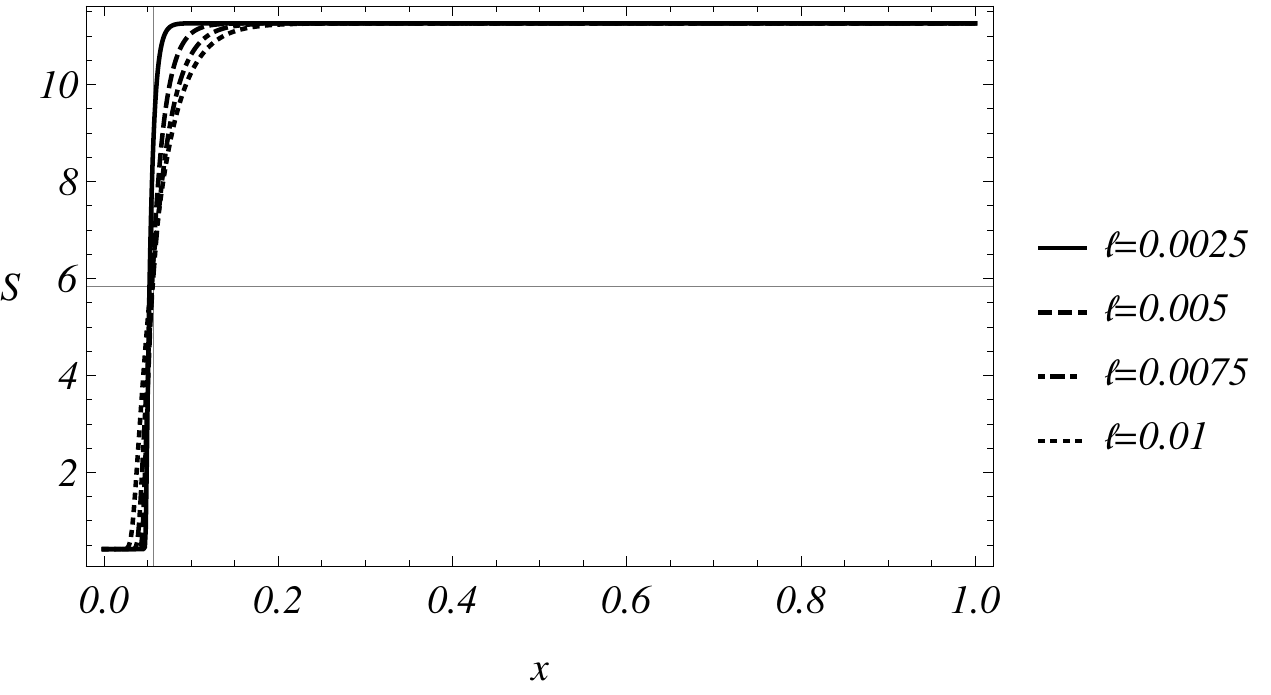}}}
}
\put(340,0)
{
\resizebox{5.75cm}{!}{\rotatebox{0}{\includegraphics{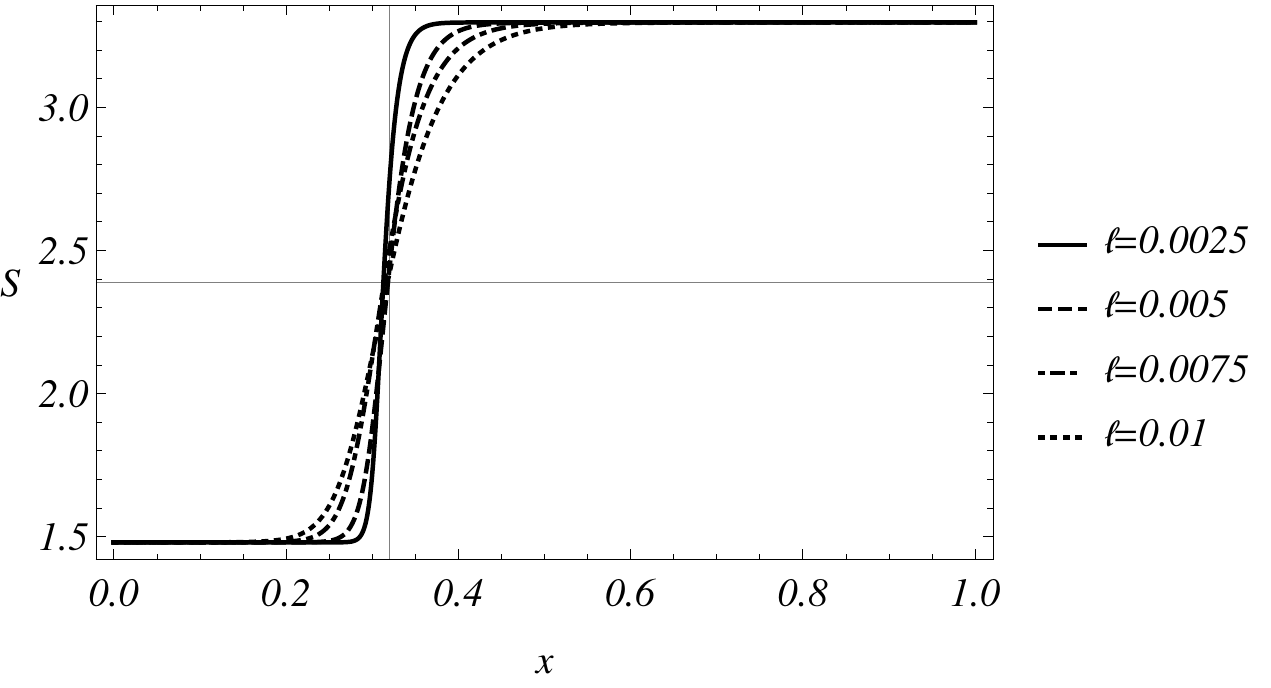}}}
}
\end{picture}
\caption{Profile \eqref{interfaccia} for the \textit{model a} with $T=283$~K and $P=33.79$~MPa (left), and $T=308$~K and $P=-1.78$~MPa (center). Profile \eqref{interfaccia} for the \textit{model b} with $T=308$~K and $P=-0.70$~MPa (right).}
\label{f:chidep080}
\end{figure*}

\begin{figure}
\centering
\includegraphics[width=0.35\textwidth]{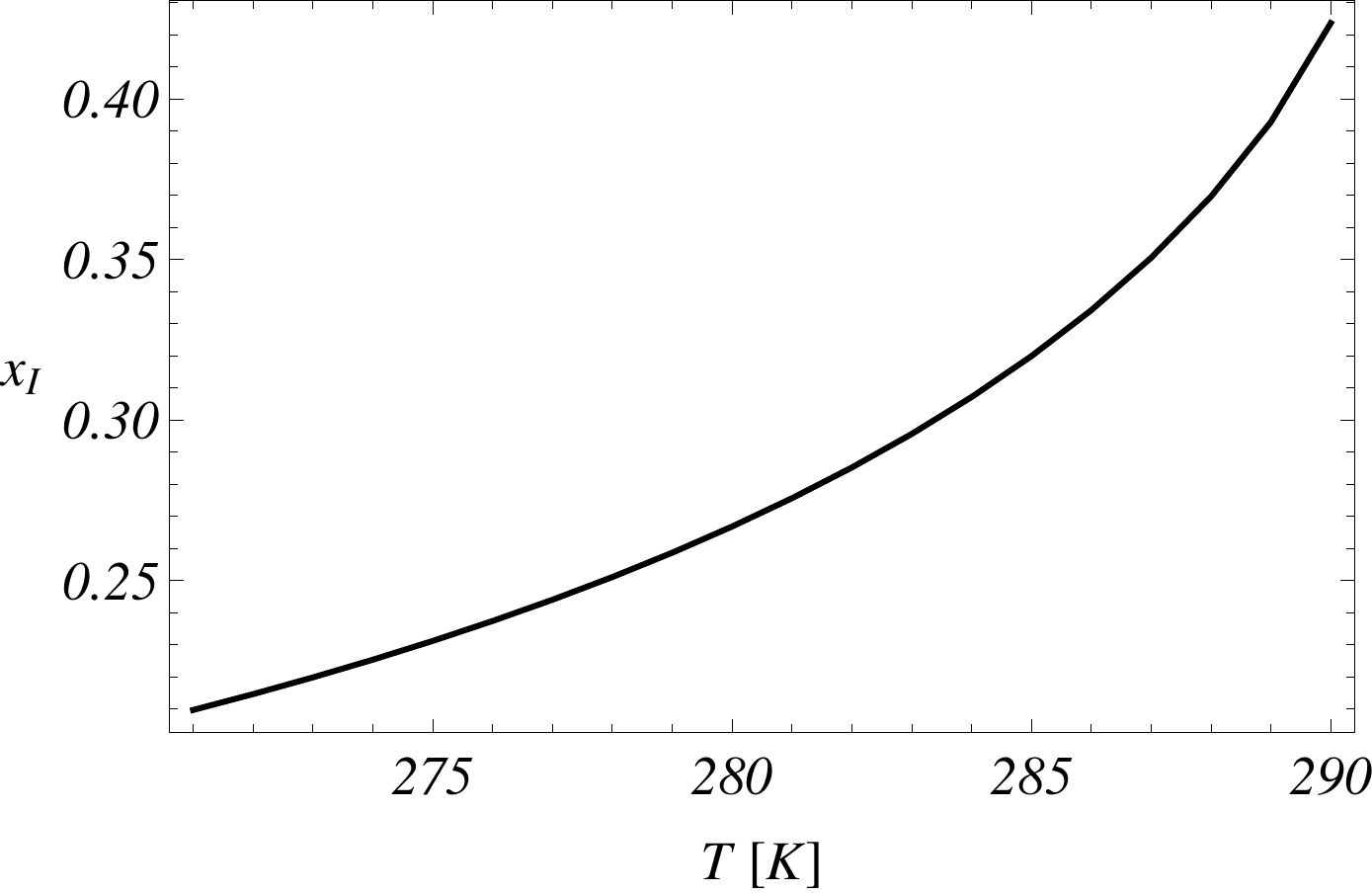}
\includegraphics[width=0.35\textwidth]{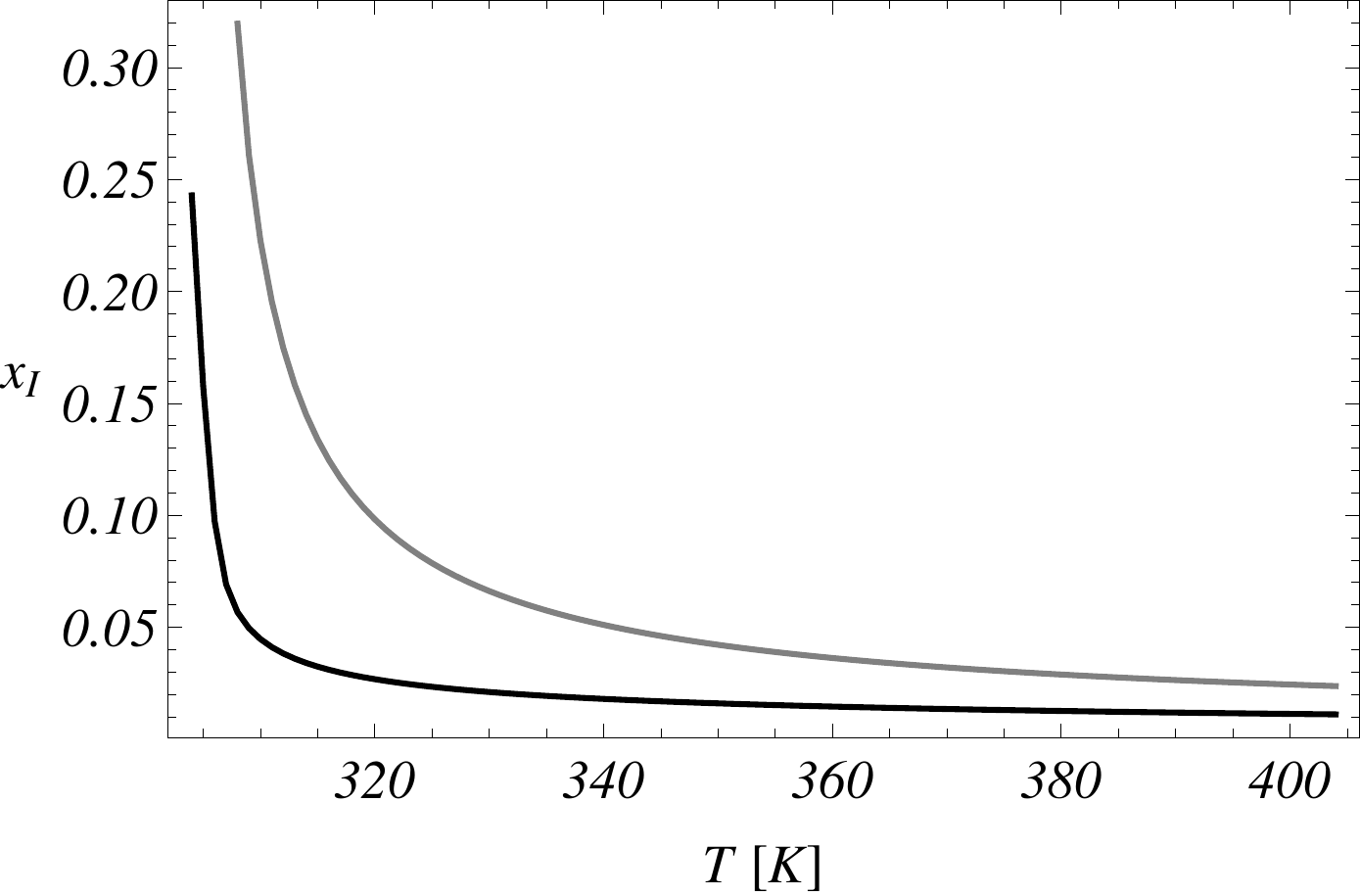}
\caption{Interface position \eqref{interfaccia} as a function of temperature at coexisting pressure. On the left results for the \textit{model a} at $T<T_{\textrm{low}}$ are plotted. On the right results for the \textit{model a} (black) and for the \textit{model b} (gray) at $T>T_{\textrm{high}}$ are plotted.}
\label{f:chidep100}
\end{figure}


In figure~\ref{f:chidep080},  we show 
that the interface tends to a sharp kink when the stiffness is chosen 
smaller and smaller. We introduce a physically more relevant parameter, 
the \textit{characteristic length} $\ell$ defined as 
\begin{equation}
\label{elle}
\ell^2=\frac{\Omega}{R T} \kappa\,.
\end{equation}
We plot the solution of the equation \eqref{Euler-Lagrange} with 
Dirichlet boundary conditions \eqref{bound} for different values of 
the characteristic length $\ell$ at given temperature and pressure 
ensuring coexistence. The plots show that the interface tends 
to localize in the small stiffness limit. 
In figure~\ref{f:chidep080} (left and central panel) we solved the problem 
for the \textit{model a}, whereas in figure~\ref{f:chidep080}
(right panel) 
the \textit{model b} was considered. 
We remark that the boundary value problem 
\eqref{Euler-Lagrange}--\eqref{bound} 
has been numerically solved by means of a properly 
implemented finite element code.

It is worth noting that, as already remarked above, due to the 
fact that even at coexistence the energy $G$ is not 
symmetric with respect to the central local maximum
the position of the interface depends on the 
temperature.
From the physical point of view, this means that at different 
temperatures the relative portions of the sample occupied 
by the shrunk and the swollen phases change. 

It is interesting to compare the center and right panels in 
figure~\ref{f:chidep080}.
At temperature $T=308$~K, 
the \textit{model a} predicts that the interface position 
is close to the boundary where the shrunk phase is preserved; 
this means that the sample is mostly occupied by the swollen 
phase. On the other hand, at the same temperature, 
the \textit{model b} predictions 
are slightly different, indeed the interface position is close 
to $0.3$ so that a not negligible part of the system is occupied 
by the shrunk phase.  

In figure~\ref{f:chidep100} the interface position 
is plotted as a function of the temperature for both models. 
The difference between the predictions of \textit{model a} and \textit{model b} is pointed out in the picture on the right. 
From both pictures we conclude that at temperatures far from the 
limiting values $T_\textrm{low}$ and $T_\textrm{high}$ the interface 
position is close to zero so that the sample is mostly occupied by 
the swollen phase. Conversely, the interface moves towards the center 
of the sample when the temperature gets closer to its limiting 
values. This effect is much more important for the \textit{model b}. 

We remark that our result goes in the opposite direction 
with respect to that illustrated in \citep[Fig.~14d]{cai2011}. 
There, indeed, it is stated that at large temperature 
the shrunk phase tends to fill up the whole sample. 
However, 
we have to notice that the problems considered in our paper and 
in \citep{cai2011} are similar, but not equivalent, indeed, 
the setup in that paper is fully three--dimensional, whereas 
our discussion is limited to dimension one.

\section{Conclusions}
We addressed the modeling of thermally--driven volume transition in hydrogels, within the Flory--Rehner thermodynamic setting; precisely, accounting for the temperature--depending pattern of the interaction parameter determined in \citep{hirotsu1989} for NIPA hydrogels, and in \citep{afroze2000} for aqueous solutions of uncrosslinked PNIPAM. In both models, the Flory parameter depends 
linearly 
on the volume fraction of polymer within the gel; on the contrary, the dependence on temperature is different. We proposed a detailed analysis aimed to establish the ranges of both temperature and traction which allow for the coexistence of two different phases (swollen and shrunk) in the hydrogel. With specific reference to a one--dimensional problem, we  showed as different models for the interaction parameter deliver different conclusions.

Finally, for the values of temperature and traction ensuring phase coexistence, we
presented a gradient model, appropriately extending the Flory--Rehner energy,
and localize the interface position between the shrunk and the swollen phase, on one hand by means of a phase space analysis, previously developed by some of the Authors, on the other developing proper finite element numerical calculations.

%
%
\end{document}